\documentclass[11pt,twoside]{article} 
\usepackage{asp2004}
\usepackage{epsf}
\usepackage{psfig}
\usepackage{lscape} 

\begin{document} 
\title{Analysis of DA White Dwarfs from the \\
McCook \& Sion Catalog}
\author{A. Gianninas, P. Bergeron, and P. Dufour} 
\affil{D\'epartement de Physique, Universit\'e de Montr\'eal, C.P. 6128, Succ. Centre-Ville, Montr\'eal, Qu\'ebec, Canada, H3C 3J7}
\begin{abstract} 
The Sloan Digital Sky Survey has already doubled the number of
spectroscopically identified white dwarfs. However, there remains a
wealth of white dwarfs in the McCook \& Sion catalog for which little
or nothing is known. We have thus undertaken a spectroscopic survey of
all Northern hemisphere DA white dwarfs from this catalog in order to
(1) confirm the spectroscopic classification, and (2) provide
measurements of $T_{\rm eff}$ and $\log g$ for each star. As part of
this project, we have also secured spectroscopic data for all DA stars
with high-speed photometric measurements available, in order to study
the purity of the ZZ Ceti instability strip, which has recently been
challenged, once again.
\end{abstract}

\section{The Survey}

High Balmer line spectroscopic observations of DA stars from the
McCook \& Sion catalog were secured using the Steward Observatory
2.3-m telescope equipped with the Boller \& Chivens
spectrograph. Objects were selected on the basis of (1) their
temperature index lying between 3 and 7, (2) having an apparent visual
magnitude brighter than 17 and (3) being situated in the Northern
hemisphere (i.e. declination greater than $-$30 degrees). Eventually, we
are planning to secure similar observations for all bright ($V <$ 17)
white dwarfs of all spectral types in the catalog. In our attempt to
provide a spectroscopic confirmation of all DA stars in the McCook \&
Sion catalog, we have found 22 stars classified DA in the McCook \&
Sion catalog that are clearly lower gravity objects. These represent
$14\%$ of the stars which we have observed to date.

\subsection{Results}

All DA spectra were fitted with the spectroscopic method described in
Bergeron et al. (1992) to determine $T_{\rm eff}$ and $\log
g$. Stellar masses are then derived using the evolutionary models of
Wood (1995) with thick hydrogen layers. In Figure 1 we show the
resulting mass distribution as a function of effective temperature for
a total of 770 stars surveyed thus far. Also superposed are isochrones
based on the evolutionary models of Wood (1995) and Fontaine et
al. (2001) at different ages along the white dwarf cooling sequence
(ages are indicated in Gyr). The solid lines correspond to the white
dwarf cooling times only, while the dotted lines include the main
sequence lifetime.

\begin{figure}[!ht]
\plotone{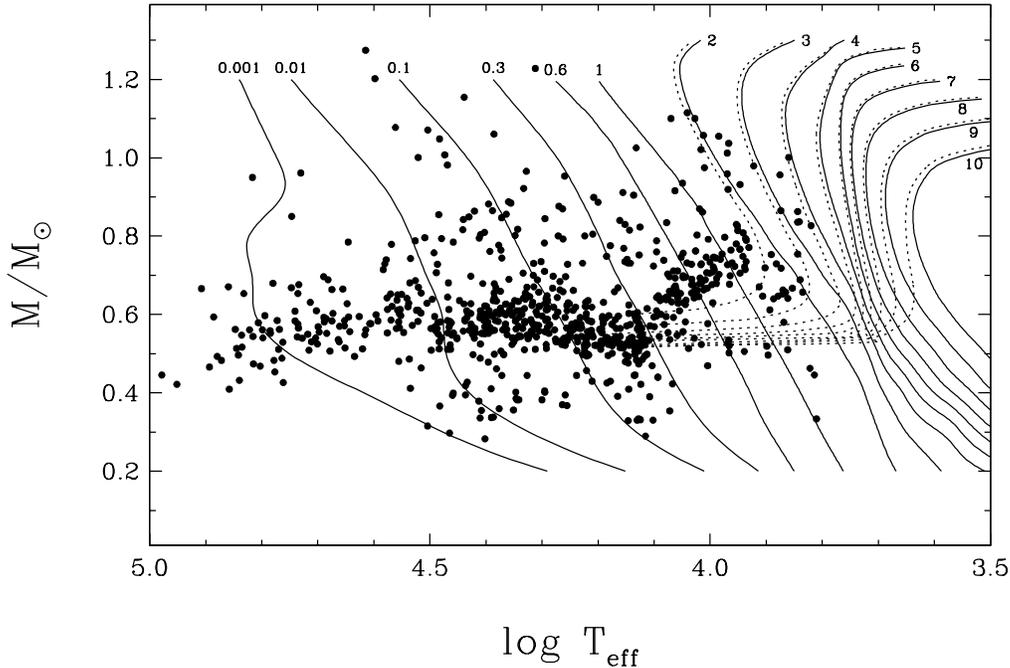}
\caption{Mass distribution of all DA white dwarfs in our sample as a function of $\log T_{\rm eff}$. The solid lines correspond to isochrones of white dwarf cooling times only, while the dotted lines include the main sequence lifetime as well.}
\end{figure}

The most striking feature here is the upturn of the sequence at $\log
T_{\rm eff}$$\sim$4.1, or $T_{\rm eff}$$\sim$13,000
K. Above this temperature, the sequence shows a distribution
consistent with an average mass near 0.6 $M_{\sun}$. Since there is no
reason, a priori, for white dwarf masses to increase as they cool
below 13,000 K, our results may indicate some inadequate treatment of
the physics of white dwarf atmospheres (convection, line broadening,
opacities, etc.). Alternatively, our basic physical assumptions may be
incorrect, such as the atmospheric composition, which is assumed to be
of pure hydrogen. Indeed, it is known that the presence of
spectroscopically invisible helium in cool DA stars may affect the
determinations of $\log g$ (see Boudreault \& Bergeron, these
proceedings).

Finally, we see that the majority of stars have inferred ages smaller
than 10 Gyr based on the isochrones that include main sequence
lifetimes. DA stars below $M\sim$0.47 $M_{\sun}$ must have followed a
different evolutionary path, most likely through binary star
evolution.

\subsection{Mass Distribution}

The mean mass of our sample turns out to be fairly high at 0.620 $\pm$ 0.140 
$M_{\sun}$. However, when the sample is divided into two
groups, above and below 13,000 K, the mean mass of the hot component
decreases to a value of 0.596 $\pm$ 0.131 $M_{\sun}$, more consistent with the
results of previous spectroscopic analyses.  The cool component has a
mean of 0.693 $\pm$ 0.143 $M_{\sun}$, significantly larger than that of hotter
stars. This peculiar yet interesting result is currently being
investigated by our group.

\section{The ZZ Ceti Instability Strip}

Figure 2 shows an enlargement of the area in the $T_{\rm eff}$-$\log
g$ plane that contains the ZZ Ceti instability strip.  The filled
circles represent the 36 ZZ Ceti stars from Bergeron et al. (2004),
while the open circles correspond to all DA stars that have been
confirmed as photometrically constant by various
investigators. Finally, the triangles represent all the remaining DA
stars in our sample that have not been investigated for photometric
variability, to our knowledge.

\begin{figure}[!ht]
\plotone{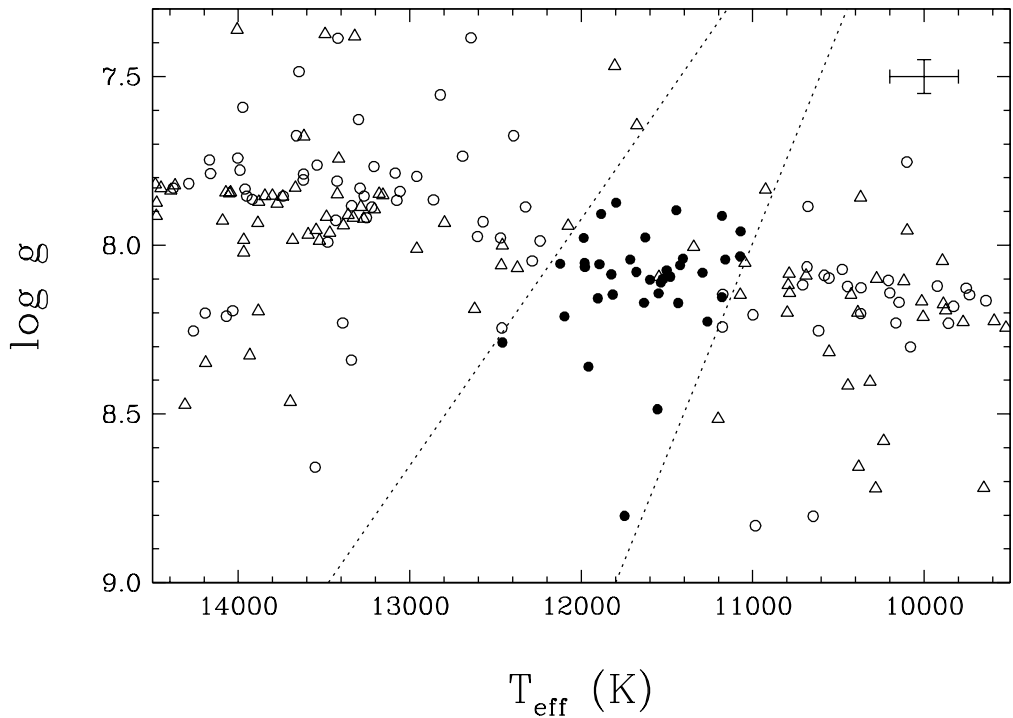}
\caption{The region of the $T_{\rm eff}$-$\log g$ plane which contains the ZZ Ceti instability strip.}
\end{figure}

If we consider the photometric samples only (open and filled circles),
our result is clearly consistent with the ZZ Ceti instability strip
being pure, with all constant stars falling outside the empirical
strip, shown as dotted lines in Figure 2. This is clearly at odds with
the recent results obtained by Mukadam et al. (2004) who claim, based
on the analysis of the DA stars in the SDSS, that the ZZ Ceti
instability strip contains a significant fraction of non-variable
stars.  We believe that high signal-to-noise (S/N $>$ 50)
spectroscopic observations are required to achieve the results shown
here, a quality that is not necessarily achieved in the SDSS data.

Finally, there are 2 program stars that fall well inside the
instability strip. One of these two, HE 1429$-$0343, has already been
confirmed as a new ZZ Ceti pulsator (see Silvotti et al., these
proceedings). With regards to the remaining candidate, high-speed
photometric observations are planned in order to ascertain its variability.

\section{GD 362: A New Metal-Rich DAZ Star}

During the course of large surveys, it is often the case that
interesting objects are uncovered. This is the case with GD 362,
previously classified as a subdwarf, or possibly as having a composite
spectrum. GD 362 is in fact a cool DAZ star, as displayed in Figure
3. This is quite an unusual object showing a number of metallic lines
in its spectrum. Also shown is our best fit to the data with
the atmospheric parameters indicated in the Figure. The major spectral
metallic features are indicated as well. A more detailed analysis of
this object will be presented elsewhere.

\begin{figure}[!ht]
\plotone{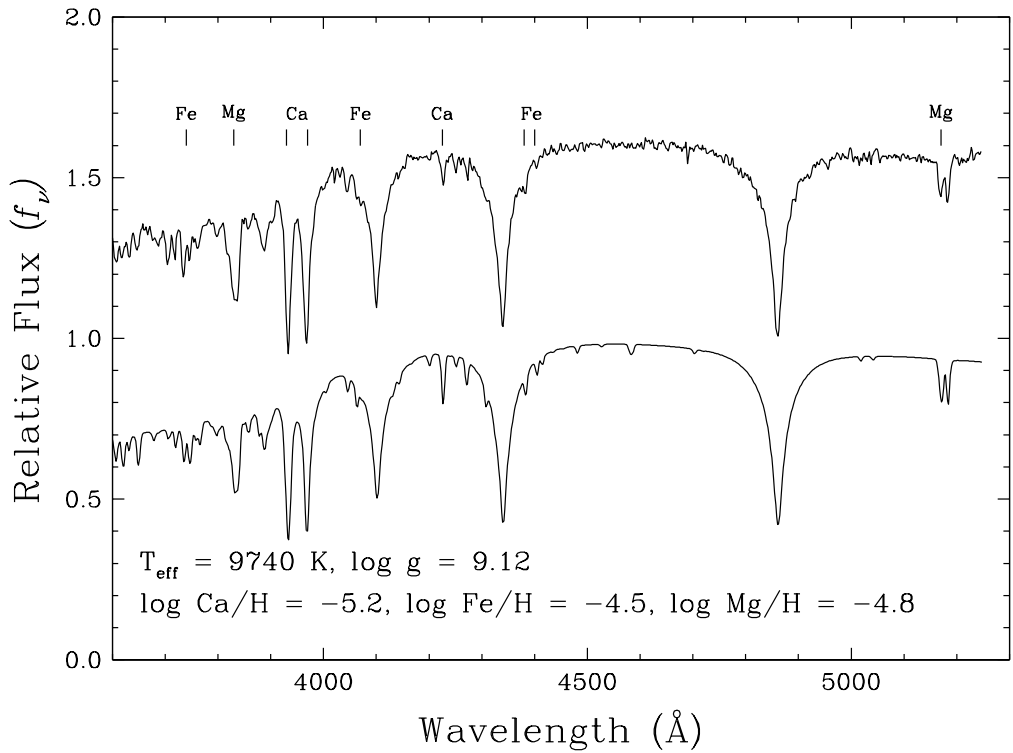}
\caption{Observed ({\it top}) and synthetic ({\it bottom}) spectra of GD 362.}
\end{figure}

\acknowledgements{This work was supported in part by the NSERC\\ Canada
and by the Fund FQRNT (Qu\'ebec)}.


\begin{references}
\reference Bergeron, P., Saffer, R. A., \& Liebert J. 1992, \apj, 394, 228
\reference Bergeron, P., Fontaine, G., Bill\`eres, M., et al. 2004,
\apj, 600, 404
\reference Fontaine, G., Brassard, P., \& Bergeron, P. 2001, \pasp, 113, 409
\reference Mukadam A. S., Winget, D. E., von Hippel, T., et al. 2004,
\apj, 612, 1052
\reference Wood, M. A., 1995, in White Dwarfs, Proc. of the 9$^{th}$
European Workshop on White Dwarfs, eds. D. Koester \& K. Werner
(Berlin: Springer), 41
\end{references}
\end{document}